\documentclass[aps,pra,twocolumn,showpacs,superscriptaddress,10pt,longbibliography]{revtex4-2}

\usepackage[colorlinks = true,linkcolor = red,citecolor = magenta]{hyperref}
\usepackage[sort&compress]{natbib}
\usepackage{scalerel}
\usepackage[normalem]{ulem}
\usepackage{xcolor}
\usepackage{bm}
\usepackage{amsmath}
\usepackage{bbm}
\usepackage{amsfonts}
\usepackage{amssymb}
\usepackage{graphicx}
\usepackage{subfigure}
\usepackage{mathtools}
\usepackage{mathrsfs}
\usepackage{physics}
\usepackage{dsfont}
\usepackage{float}
\usepackage{lmodern}
\usepackage[many]{tcolorbox}
\usepackage{empheq}
\usepackage[capitalise]{cleveref}
\usepackage{textcomp}
\usepackage[dvipsnames]{xcolor}

\newcommand{\scs}{\scriptscriptstyle}

\begin{document}

\title{Current statistics for quantum transport along a bosonic tight-binding chain connected to Markovian reservoirs}
\author{P. S. Muraev}
\affiliation{Kirensky Institute of Physics, Federal Research Centre KSC SB RAS, 660036, Krasnoyarsk, Russia}
\affiliation{IRC SQC, Siberian Federal University, 660041, Krasnoyarsk, Russia}
%\affiliation{School of Engineering Physics and %Radio Electronics, Siberian Federal University, %660041, Krasnoyarsk, Russia}
\author{A. R. Kolovsky}
\affiliation{Kirensky Institute of Physics, Federal Research Centre KSC SB RAS, 660036, Krasnoyarsk, Russia}
\affiliation{IRC SQC, Siberian Federal University, 660041, Krasnoyarsk, Russia}
\affiliation{School of Engineering Physics and Radio Electronics, Siberian Federal University, 660041, Krasnoyarsk, Russia}
\author{D. N. Maksimov}
\affiliation{Kirensky Institute of Physics, Federal Research Centre KSC SB RAS, 660036, Krasnoyarsk, Russia}
\affiliation{IRC SQC, Siberian Federal University, 660041, Krasnoyarsk, Russia}

\date{\today}

\begin{abstract}
We study quantum transport of non-interacting particles in a bosonic tight-binding chain coupled to Markovian reservoirs at its boundaries. Using the analytic solution in the form of an anti-normally ordered quantum characteristic function, we obtain exact probability distribution functions for the current on each bond in the chain, as well as the inter-bond current correlation functions. We extend the analysis to tight-binding chains subject to uniform on-site phase damping. In the phase-damped case, we find that a Gaussian approximation for the quantum characteristic functions predicts current fluctuations with good accuracy.
\end{abstract}
\maketitle

\section{Introduction}
In recent years, considerable attention has been paid to the problem of nonequilibrium transport in open quantum systems. This problem is a central topic at the intersection of quantum optics, condensed matter physics, and quantum thermodynamics~\cite{Breuer2002, Esposito_2009, krinner2017, bertini2021, landi2022, kolovsky2026}. The theoretical background of quantum transport trace back to the Landauer formalism and the nonequilibrium Green's function approach, which remain standard tools and are thoroughly presented in~\cite{jauho1994, Datta1997}. The study of quantum transport frequently employs a minimal setup, namely two particle reservoirs connected by a conductor, which can range from a single quantum point contact or a tunneling junction to an extended mesoscopic structure of arbitrary geometry. Such geometries, often referred to as boundary-driven systems~\cite{lacerda2021, landi2022, kolovsky2026}, have become the basis of numerous theoretical and experimental investigations.

On the experimental side, ultracold atomic gases have proven to be a remarkably versatile platform for investigating mesoscopic transport phenomena. Two-terminal configurations with cold fermions have allowed the observation of Ohmic conduction~\cite{brantut2012}, quantized conductance~\cite{krinner2015}, and spin-selective or dissipative point contacts~\cite{lebrat2019, corman2019}. Superfluidity and strong correlations open up avenues for investigating physical effects such as resistance drop at superfluid transition~\cite{stadler2012}, interaction-assisted reversal of thermopower~\cite{hausler2021}, enhanced irreversible entropy transport~\cite{fabritius2024}, and transport through a dark state in a unitary superfluid~\cite{talebi2024}. Mesoscopic lattices with individually controlled sites further enable the study of band and correlated insulating phases~\cite{lebrat2018}. Beyond cold atoms, quantum simulation platforms based on transmons~\cite{fedorov2021, zhang2024} have emerged as powerful tools for studying steady-state transport.

In parallel, extensive theoretical efforts have been focused on transport through extended mesoscopic structures of arbitrary geometry, such as chains, ladders, and lattices. These efforts have covered bosonic~\cite{Asadian2013, Manzano2016, kolovsky2018, bychek2020, muraev2021, muraev2022} and fermionic~\cite{Ajisaka2012, 120, maksimov2024, muraev2025} systems, numerical and analytic solutions for current in %the ${\rm XX}$~chain 
two-level chains with dephasing~\cite{znidaric10, Manzano2012, lacerda2021, Znidaric2024}. Further important aspects include transport between finite reservoirs~\cite{amato2020,Xu2022}, local particle loss~\cite{uchino2022}, full counting statistics with engineered gain and loss channels~\cite{ganguly2026}, and transport in complex networks~\cite{Witt2013, Manzano2013, Kurt2023, Saha2026}.
A separate line of research concerns transport through a periodically driven quantum point contact connecting two fermionic chains~\cite{Gamayun2021, Ermakov2023, Dudinets2025}. Beyond specific models, master-equation and finite-time Landauer approaches~\cite{nietner2014, Gruss_2016} as well as Liouville–von Neumann techniques for molecular junctions~\cite{zelovich2014, zelovich2016} have broadened the theoretical toolkit. {Field-theoretical approaches on the Keldysh contour, applied to quadratic Liouvillians in the coherent-state basis, have enabled the derivation of generalized transport formulas and quantum kinetic equations for open systems with multiple dissipative channels~\cite{aksenov2026}.

%}

A particularly illustrative example of quantum transport is a tight-binding bosonic chain connected at its ends to Markovian reservoirs, where a difference between the mean particle densities in the reservoirs drives a stationary current through the chain. Uniform on-site dephasing, which suppresses phase coherence while preserving the global symmetry $\rm U(1)$, is a natural addition when one wishes to understand how transport survives the loss of coherence. The problem is relevant both for the theory of nonequilibrium steady-states and for the analysis of current fluctuations in quantum transport. A standard approach to describing such setups is provided by the Lindblad master equation. For boundary-driven tight-binding chains, this approach was used, for example, in~\cite{Asadian2013, Manzano2016, Guo2017, kolovsky2018, jin2020, bychek2020}. The analytic solution for current in the bosonic chain with dephasing was obtained in ~\cite{Asadian2013} where the authors showed that in the absence of dephasing the system exhibits ballsitic transport whereas with non-zero dephasing rate the conductance drops as $1/L$ with $L$ as the number of cites in the chain.
These results were extended to higher dimension in ~\cite{Manzano2016}. An analytic framework for current in quadratic open bosonic and fermionic systems was proposed in~\cite{Guo2017}. For generic systems with Lindblad boundaries, a compact transport formula was obtained in ~\cite{jin2020} using the Keldysh path‑integral formalism.

For non-interacting carriers the mean currents and site populations can be obtained by solving the master equation for the single‑particle density matrix (SPDM). However, complete characterization of transport must also address \textit{current fluctuations}. One particularly powerful framework for this purpose is full counting statistics (FCS), pioneered by Levitov and Lesovik for noninteracting electrons~\cite{levitov1993,levitov1996}. Over the past decades, FCS has been generalized far beyond its original context. FCS is now routinely applied to bosonic systems, phonon and photon transport, and quantum master equations~\cite{Esposito_2009,Landi2024}. The central object of FCS is the characteristic function of the transferred particle number, whose logarithm yields the cumulant
generating function and thereby provides access to all cumulants, while the complete probability distribution is obtained via inverse Fourier transformation. The same idea of encoding the full statistics in a characteristic function is provided by the phase‑space formalism~\cite{gardiner2004,Gardiner2014}.

\begin{figure}
    \centering
    \includegraphics[width=1.0\linewidth, height=0.2\textwidth]{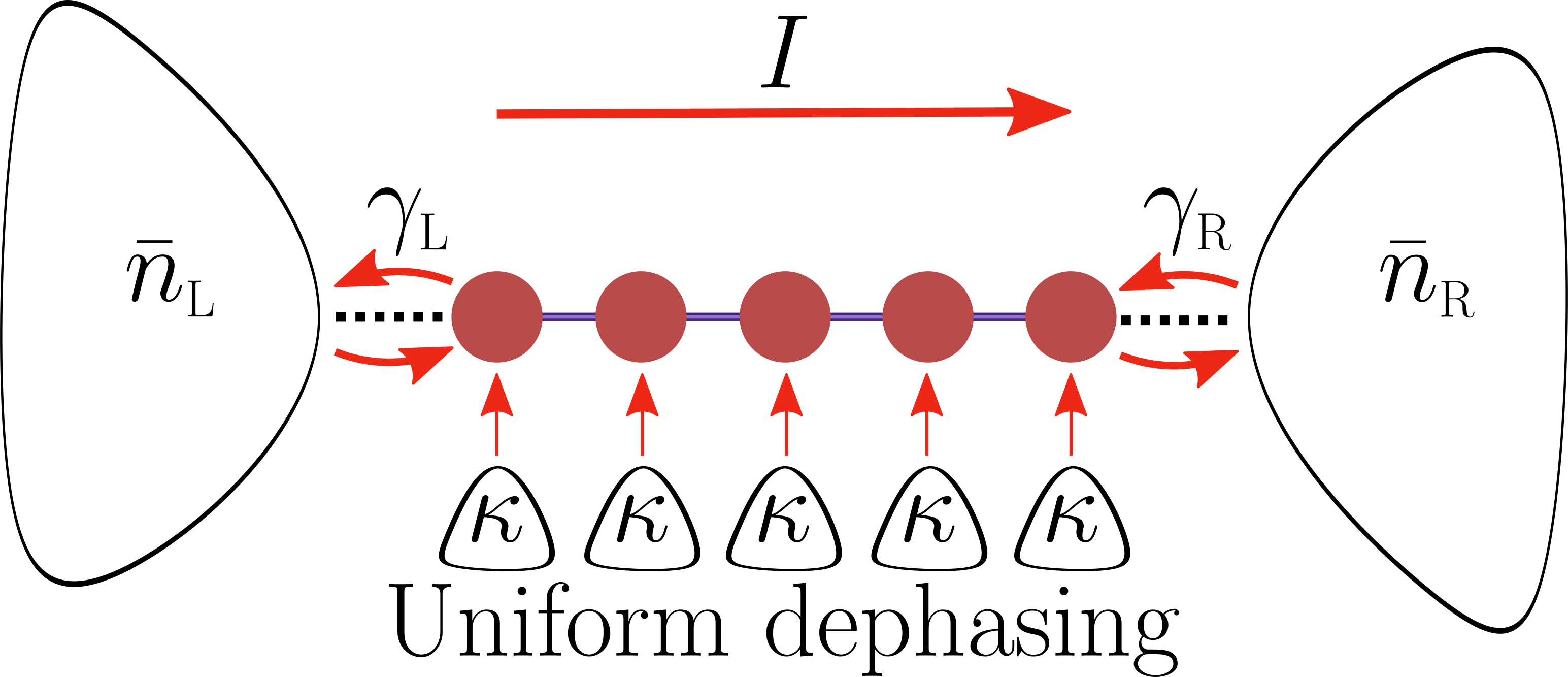}
    \caption{Particle transport through a tight-binding chain coupled to Markovian reservoirs. The current $I$ flows from the left to the right driven by the mean particle density imbalance $\bar{n}_{\scs{\rm L}}-\bar{n}_{\scs{\rm R}}$. Relaxation rates $\gamma_{\scs{\rm L}}$ and $\gamma_{\scs{\rm R}}$ correspond to the particle exchange rates between each reservoir and the coupled site of the chain.}
    \label{fig:scheme}
\end{figure}

In this work, we perform a comprehensive analysis of a bosonic tight-binding chain with Markovian reservoirs at both ends and uniform on-site dephasing, see Fig.~\ref{fig:scheme}. Our approach combines the Lindblad master equation with phase-space techniques. For quadratic Hamiltonians and jump operators linear in the field operators, i.e. for quasi-free systems~\cite{barthel2022, Teretenkov2019}, the steady-state is Gaussian. In this case, which corresponds to zero dephasing rate, the characteristic function is fully determined by the SPDM, making it possible to compute not only mean currents and density profiles but higher moments of the distribution. In what follows, we adopt the anti‑normally ordered convention, in which the characteristic function is the Fourier transform of the Husimi $Q$‑function, and obtain the analytic solution in a closed form. The solution is then applied for deriving the current distribution functions. In the presence of dephasing the steady-state solution is not Gaussian since the jump operator is quadratic in terms of creation and annihilation operators. This significantly complicates the analysis of current fluctuations ~\cite{Ishiyama2025,Ishiyama2025Current}. In particular, in ~\cite{Ishiyama2025} the exact time-dependent density profile for domain-wall and alternating initial states was obtained by means of the Bethe Ansatz, whereas deriving the FCS of the integrated current in ~\cite{Ishiyama2025Current} required a mapping to the one-dimensional Hubbard model with imaginary interaction. Here we take a different route by applying a Gaussian approximation based on the exact solution for the SPDM. We demonstrate that the approximate solutions for the current distribution functions are in good agreement with the results of numerical simulations.

\section{The model} \label{sec:II}
We consider a tight-binding chain, which coupled to Markovian reservoirs at both ends
%~\cite{kolovsky2018,bychek2020} 
and subject to uniform dephasing along the chain, see Fig.~\ref{fig:scheme}. The Hamiltonian is given by 
\begin{equation}\label{Hamilton_ch}
    \widehat{\cal H} =-\frac{J}{2}\sum_{\ell = 1}^{L-1}\Bigl(\hat{a}^{\dagger}_{\ell+1}\hat{a}_{\ell} + \mathrm{h.c.}\Bigr),
\end{equation}
where $\hat{a}_{\ell}^{\dagger}$ and $\hat{a}_{\ell}$ are bosonic creation and annihilation operators satisfying the canonical commutation relation  $[\hat{a}_{\ell},\hat{a}^{\dagger}_{\ell '}]=\delta_{\ell,\ell'}$, $J$ is the hopping amplitude, and $L$ is the chain length. We describe the state of the chain by using the reduced density matrix $\widehat{{\cal R}}$. The coupling to Markovian reservoirs and dephasing baths is accounted for by the Lindblad master equation
\begin{equation}\label{m_eq}
    \frac{\partial \widehat{{\cal R}}}{\partial t}=-i[\widehat{{\cal H}}, \widehat{{\cal R}}]+
    \sum_{\ell=1,L}{\cal L}_{\ell}(\widehat{\cal R}) +  \sum_{\ell=1}^L{\cal D}_{\ell}(\widehat{\cal R}),
\end{equation}
where ${\cal L}_{\ell}$ are Lindblad superoperators or dissipators that describe particle exchange between the chain's edge sites and the left, $\ell=1$, or right, $\ell=L$, reservoir. The explicit form of these superoperators is
\begin{align}\label{eq_supop}
    \begin{split}
        {\cal L}_{\ell}(\widehat{\cal R}) &= \gamma_{\ell}(\bar{n}_{\ell}+1)\left(\hat{a}_{\ell}\widehat{{\cal R}}\hat{a}_{\ell}^{\dagger} - \frac{1}{2}\{\hat{a}_{\ell}^{\dagger}\hat{a}_{\ell},\widehat{{\cal R}}\}\right) \\
        &+\gamma_{\ell}\bar{n}_{\ell}\left(\hat{a}_{\ell}^{\dagger}\widehat{{\cal R}}\hat{a}_{\ell} - \frac{1}{2}\{\hat{a}_{\ell}\hat{a}_{\ell}^{\dagger},\widehat{{\cal R}}\}\right),
    \end{split}
\end{align}
 where $\{A,B\}=AB + BA$ denotes the anticommutator, $\gamma_{\ell}$ is the relaxation rate, and $\bar{n}_{\ell}$ is the mean density of particles in the corresponding reservoir. 
To be consistent with the definitions in Fig.~\ref{fig:scheme} we assume in the index notations that $1={\scs{\rm L}}$ and $L=\scs{{\rm R}}$ when applied to the quantities $\bar{n}_{\ell}$ and $\gamma_{\ell}$.
 The dephasing superoperator has the following form
\begin{equation}\label{deph_super}
{\mathcal{D}}_{\ell}(\widehat{\mathcal{R}})
= \kappa\Bigl(\hat{n}_{\ell}\widehat{\mathcal{R}}\hat{n}_{\ell}
- \frac{1}{2}\{\hat{n}_{\ell}^{2},\widehat{\mathcal{R}}\}\Bigr),
\end{equation}
where \(\kappa\) is the dephasing rate and $\hat{n}_{\ell}=\hat{a}_{\ell}^{\dagger}\hat{a}_{\ell}$.

% \section{steady-state single‑particle density matrix}\label{sec:III}
% While the phase‑space formalism gives the full counting statistics, it requires solving a multi‑dimensional partial differential equation (PDE) for the characteristic function. 

For many physically relevant quantities, such as the site populations and the average current, it is sufficient to know only the first and second moments of the field operators. In our model, the Hamiltonian is quadratic and the jump operators are either linear (reservoir terms) or quadratic (dephasing) in the creation/annihilation operators. 
As we shall see, the equations of motion for the SPDM close as they only involve the SPDM itself.
Thus, we can derive a closed linear system for the SPDM which is quite simple to handle.
The SPDM $\hat{\rho}$ is defined through its elements as~\cite{maksimov2024,rosati2014,giesbertz2019}
\begin{equation}\label{eq:SPDM_def}
  \rho_{\ell,\ell'}\!:=\langle\hat a_{\ell'}^\dagger\hat a_{\ell}\rangle = {\rm Tr}\left[\hat{a}^{\dagger}_{\ell'}\hat{a}_{\ell}\widehat{\cal R}\right].  
\end{equation}
We also mention that some authors use the transposed convention $\rho_{\ell,\ell'}= \langle\hat{a}^{\dagger}_{\ell}\hat{a}_{\ell'}\rangle$, see~\cite{Witthaut2011,kolovsky2018,amato2020,bychek2020,muraev2021,muraev2022}, which yields the same physical observables after a suitable exchange of indices. We adopt Eq.~\eqref{eq:SPDM_def} throughout this work for a reason to be explained below.

In the Fock space the current between the $\ell$ and $\ell+1$ sites is calculated with the formula
\begin{equation}
     \bar{I}_\ell={\rm Tr} (\widehat{\cal I}_{\ell}{\widehat{\cal R}}),
\end{equation}
where $\widehat{\cal I}_{\ell}$ is the current 
operator\begin{equation}\label{cur_op}
    \widehat{\cal I}_{\ell} = \frac{J}{2i}\left(\hat{a}^{\dagger}_{\ell}\hat{a}_{\ell+1} - \mathrm{h.c.}\right).
\end{equation} 
Equation~\eqref{cur_op} leads to the following formula in terms of the SPDM
\begin{equation}\label{I_SPDM}
    \bar{I}_\ell
    = \frac{J}{2i}\bigl(\rho_{\ell+1,\ell}-\rho_{\ell,\ell+1}  \bigr).
\end{equation}
Therefore, the current operator in the Wannier basis, such that $\bar{I}_{\ell}={\rm Tr} (\hat{I}_{\ell}\hat{\rho})$, takes the following form
\begin{equation}\label{cur_op_SPDM}
    \hat{I}_{\ell}=\frac{J}{2i}\left(|\ell\rangle\langle\ell +1| - {\rm h.c.}\right),
\end{equation}
where $|\ell\rangle$ is the $\ell$ site Wannier state.
The mean currents between the reservoirs and the edge sites can be calculated by applying the expression below
\begin{align}\label{cur_res}
    \begin{split}
        \bar{I}_{\scs{\rm L}} &= \gamma_{\scs{\rm L}}(\bar{n}_{\scs{\rm L}} - \langle\hat{n}_{1}\rangle), \\
         \bar{I}_{\scs{\rm R}} &= \gamma_{\scs{\rm R}}(\langle\hat{n}_{\scs{L}}\rangle - \bar{n}_{\scs{\rm R}}).
    \end{split}
\end{align}
where $\langle \hat{n}_{1}\rangle =\rho_{1,1}$ and $\langle \hat{n}_{\scs{L}}\rangle =\rho_{\scs{L},\scs{L}}$ are the mean occupation numbers for the first and last lattice sites, respectively.

The equation of motion for the SPDM can be derived directly from the master equation~\eqref{m_eq}.  It takes the form
\begin{align}\label{eq:SPDM}
\begin{split}
    \frac{\partial{\hat{\rho}}}{\partial t} &= -i(\widehat{H}_{\mathrm{\scs{eff}}}\hat{\rho}-\hat{\rho}\widehat{H}_{\mathrm{\scs{eff}}}^{\dagger})  \\
    +\kappa&\sum_{\ell=1}^L|\ell\rangle\langle \ell| \hat{\rho}| \ell\rangle\langle \ell|
    +\sum_{\ell= { 1,L}}\gamma_{\ell}\bar{n}_{\ell}|\ell\rangle\langle \ell|  
\end{split}
\end{align}
with the effective non‑Hermitian Hamiltonian defined as follows
\begin{equation}\label{eq:Heff}
\begin{split}
    \widehat H_{\mathrm{\scs{eff}}} &= -\frac{J}{2}\sum_{\ell=1}^{L-1}
    \bigl( |\ell+1\rangle\langle\ell| + \mathrm{h.c.} \bigr) \\
    &- \frac{i}{2}\sum_{\ell=1,L }\gamma_{\ell}\, |\ell\rangle\langle \ell|
    - i\frac{\kappa}{2}\sum_{\ell=1}^L |\ell\rangle\langle\ell|.    
\end{split}
\end{equation}
Note that Eq.~\eqref{eq:SPDM} is not of the Lidblad form since the trace of the SPDM is not preserved due to particle exchange.
We mention in passing that the transposed definition of the SPDM would result in a change of sign $J\rightarrow -J$ in Eq.~\eqref{eq:Heff}. Therefore, we opted for reversed index ordering in Eq.~\eqref{eq:SPDM_def}. 

In the steady-state, $\partial \hat\rho/\partial t=0$, Eq.~\eqref{eq:SPDM} becomes a linear algebraic system for the matrix elements of the stationary SPDM, $\hat\rho_{\rm st}$.  Because the chain is one‑dimensional and only nearest‑neighbor couplings appear, the solution can be found in the form of a tridiagonal matrix
\begin{equation}\label{eq:SPDM_form}
\hat\rho_{\rm \scs{st}} =
\begin{pmatrix}
C & -iB & & & 0 \\
iB & A_2 & -iB & & \\
& \ddots & \ddots & \ddots & \\
& & iB & A_{L-1} & -iB \\
0 & & & iB & D
\end{pmatrix},
\end{equation}
where $A_{\ell}=\langle \hat{n}_{\ell}\rangle$ is the population of each inner site, $C=\langle \hat{n}_{1}\rangle$ and $D=\langle \hat{n}_{L}\rangle$ are the populations of the edge sites. The off‑diagonal element $B$ is the same for all bonds, which guarantees a uniform stationary current $I_{\rm st} = JB$. 
Substituting this Ansatz into the equation for the stationary SPDM 
%equation yields recurrence relations
% \begin{equation}\label{equations_SPDM}
%     \begin{split}
%         &\kappa B + \frac{\gamma_{\scs{\rm L}}\delta_{\ell,1} + \gamma_{\scs{\rm R}}\delta_{\ell +1,L}}{2} B = \frac{J}{2}(A_{\ell} - A_{\ell + 1}) \;, \\
%         &\gamma_{\scs{\rm L}} C + J B = \gamma_{\scs{\rm L}} \bar{n}_{\scs{\rm L}} \;,\\
%         &\gamma_{\scs{\rm R}} D - J B = \gamma_{\scs{\rm R}} \bar{n}_{\scs{\rm R}} \;,
%     \end{split}
% \end{equation}
% where $\ell = 1, \dots, L-1$, $A_1 = C$ and $A_{{L}} = D$. The above equations form a closed linear system for unknowns $B$, $C$, $D$, and $A_\ell$. Solving Eq.~\eqref{equations_SPDM} by forward substitution yields the stationary current
leads for the analytic solutions, first found by Asadian and co-workers~\cite{Asadian2013}, for the current
\begin{equation}\label{eq:I_deph}
I_{\rm st}
= \frac{\gamma_{\scs{\rm L}}\gamma_{\scs{\rm R}}J^2\,
        (\bar n_{\scs{\rm L}} - \bar n_{\scs{\rm R}})}
      {(\gamma_{\scs{\rm L}}+\gamma_{\scs{\rm R}})
       (\gamma_{\scs{\rm L}}\gamma_{\scs{\rm R}}+J^2)
       + 2\kappa\gamma_{\scs{\rm L}}\gamma_{\scs{\rm R}}(L-1)} ,
\end{equation}
the edge populations 
\begin{align}
\langle\hat n_1\rangle &= \bar n_{\scs{\rm L}} - I_{\rm st}/\gamma_{\scs{\rm L}}, \nonumber\\
\langle\hat n_{\scs{L}}\rangle &= \bar n_{\scs{\rm R}} + I_{\rm st}/\gamma_{\scs{\rm R}},
\end{align}
and the inner‑site populations
\begin{align}\label{eq:nj_deph}
    \begin{split}
        \langle\hat{n}_\ell\rangle &= \frac{\bar{n}_{\scs{\rm L}}+\bar{n}_{\scs{\rm R}}}{2} + \frac{I_{\rm st}}{2}\left( \frac{1}{\gamma_{\scs{\rm R}}} - \frac{1}{\gamma_{\scs{\rm L}}} \right) \\
        &+\frac{I_{\rm st}}{J^2}\left[ \kappa(L-2\ell+1) + \frac{\gamma_{\scs{\rm R}} - \gamma_{\scs{\rm L}}}{2} \right],
    \end{split}
\end{align}
where $\ell = 2,\dots, L-1$. 
% When $\kappa=0$ and $\gamma_{\scs{\rm L}}=\gamma_{\scs{\rm R}}\equiv\gamma$ these formulas reduce to the known results for a boundary‑driven tight‑binding chain without dephasing~\cite{kolovsky2018,bychek2020}. 
%This solution
%We also mention that similar formulas were obtained for the XX model with dephasing~\cite{znidaric10}. 
It is noteworthy that in the absence of dephasing, the current is independent of the length of the chain. If $\kappa\neq 0$, the transport becomes diffusive with resistance to the bias $\bar n_{\scs{\rm L}} - \bar n_{\scs{\rm R}}$ proportional to the chain length in the limit $L\rightarrow \infty$.

\section{Gaussian state and current statistics}

\subsection{Characteristic function}

While the SPDM fully determines the mean currents and occupations for any dephasing strength, the higher‑order statistics requires knowledge of the full state. In the quasi‑free limit $\kappa=0$ the stationary state is Gaussian~\cite{barthel2022}, allowing us to express high-order momenta through the SPDM.
To do that, we employ the phase-space approach. We use the anti‑normally ordered characteristic function~\cite{Gardiner2014}, which is defined as
\begin{equation}\label{def:chi}
\chi(\bm{\alpha},\bm{\alpha}^*; t):
= \operatorname{Tr}\!\Bigl[
e^{-\bm{\alpha}^\dagger  \hat{\mathbf{a}}}\,
e^{\hat{\mathbf{a}}^\dagger \bm{\alpha}}\,
\widehat{\mathcal{R}}(t)
\Bigr]. 
\end{equation}
Here \(\bm{\alpha}= (\alpha_1,\dots,\alpha_{\scs{L}})^{\intercal}\) and \(\bm{\alpha}^\dagger = (\alpha_1^*,\dots,\alpha_{\scs{L}}^*)\) are vectors of complex phase‑space variables, \(\hat{\mathbf{a}} = (\hat{a}_1,\dots,\hat{a}_{\scs{L}})^{\intercal}\) and \(\hat{\mathbf{a}}^\dagger = (\hat{a}_1^\dagger,\dots,\hat{a}_{\scs{L}}^\dagger)\) are vectors of creation and annihilation  operators, and the product is understood as the standard matrix product, e.g.,
\begin{equation}
\bm{\alpha}^\dagger\hat{\mathbf{a}} = \sum\limits_{\ell=1}^L \alpha_{\ell}^*\hat{a}_{\ell}.
\end{equation}
The characteristic function serves as a generating function of momenta. The expectation values of any anti‑normally ordered product of creation and annihilation operators are obtained by simple differentiation.  For a system of $L$ sites this reads
\begin{equation}\label{eq:moments}
    \begin{split}
        &\langle \hat{a}_1^{n_1}\cdots \hat{a}_{\scs{L}}^{n_{\scs{L}}}
        (\hat{a}_1^\dagger)^{m_1}\cdots (\hat{a}_{\scs{L}}^\dagger)^{m_{\scs{L}}} \rangle
        = \\
        &\Bigg[ \prod_{j=1}^L
        \Bigl( -\frac{\partial}{\partial\alpha_j^*} \Bigr)^{\!n_j}\,
        \Bigl( \frac{\partial}{\partial\alpha_j} \Bigr)^{\!m_j}
        \Bigg] \chi(\bm{\alpha},\bm{\alpha}^*) \Bigg|_{\bm{\alpha}=0},
    \end{split}
\end{equation}
where the product of commuting differential operators can be taken in any order. 

Applying the definition~\eqref{def:chi} to the master equation~\eqref{m_eq} yields a
linear partial differential equation for the charcteristi function
\begin{widetext}
\begin{align}\label{eq:chi_eq}
\frac{\partial\chi}{\partial t} &= \frac{iJ}{2}\sum_{\ell=1}^{L-1}
\Bigl(
- \alpha_{\ell+1}\frac{\partial\chi}{\partial\alpha_\ell}
- \alpha_\ell\frac{\partial\chi}{\partial\alpha_{\ell+1}}
+ \alpha_{\ell+1}^*\frac{\partial\chi}{\partial\alpha_\ell^*}
+ \alpha_\ell^*\frac{\partial\chi}{\partial\alpha_{\ell+1}^*}
\Bigr) 
-\sum_{{\ell}=1,L}
\Bigl[
\frac{\gamma_{\ell}}{2}
\Bigl( \alpha_{\ell}\frac{\partial\chi}{\partial\alpha_{\ell}}
      +\alpha_{\ell}^*\frac{\partial\chi}{\partial\alpha_{\ell}^*} \Bigr)
+ \gamma_{\ell}(\bar n_{\ell}+1)\alpha_{\ell}^*\alpha_{\ell}\,\chi
\Bigr] \nonumber\\
&-\frac{\kappa}{2}\sum_{\ell=1}^L\Bigl(
\alpha_\ell^2\frac{\partial^2 \chi}{\partial\alpha_\ell^2} + 
(\alpha_\ell^*)^2\frac{\partial^2 \chi}{\partial (\alpha_\ell^*)^2} -
2|\alpha_\ell|^2\frac{\partial^2 \chi}{\partial\alpha_\ell\,\partial\alpha_\ell^*} + 
\alpha_\ell\frac{\partial \chi}{\partial\alpha_\ell} + 
\alpha_\ell^*\frac{\partial \chi}{\partial\alpha_\ell^*}
\Bigr).
\end{align}
\end{widetext}
This equation is first‑order in time, but contains second‑order derivatives emerging from the dephasing superoperator.  For $\kappa=0$ Eq.~\eqref{eq:chi_eq} reduces to a first‑order linear PDE, while for $\kappa\neq0$ it is of parabolic type. Following the classification of Barthel and Zhang~\cite{barthel2022}, a bosonic system with a quadratic Hamiltonian and jump operators that are \textit{linear} in field operators is called \textit{quasi‑free}. Our model without dephasing, $\kappa=0$, belongs exactly to this class, for which the steady-state is Gaussian. Since the global symmetry $\rm{U}(1)$ forces all moments $\langle\hat a_\ell\rangle$ and $\langle\hat{a}_{\ell} \hat{a}_{\ell'}\rangle = 0$ to vanish, the characteristic function simplifies to
\begin{equation}\label{chi_gauss}
\chi(\bm\alpha,\bm\alpha^*) = \exp\bigl(-\bm\alpha^\dagger \hat M \bm\alpha\bigr),
\end{equation}
$\hat M$ is a Hermitian matrix.
Substituting the Ansatz~\eqref{chi_gauss} into Eq.~\eqref{eq:chi_eq} with $\kappa=0$ results in a closed algebraic equation for $\hat M$ that yields
\begin{equation}\label{ch_fun_sol}
\hat{M} = \hat\rho_{\scs{\rm st}} + \hat{\mathbbm{1}},
\end{equation}
where $\hat{\mathbbm{1}}$ is the identity matrix. For $\kappa=0$, 
all inner-site populations in Eq.~\eqref{eq:SPDM_form} are the same
$A_{\ell}=A$,
\begin{equation}\label{m_cur2}
   A = \frac{\bar{n}_{\scs{\rm L}}+\bar{n}_{\scs{\rm R}}}{2} + 
   \frac{(\gamma_{\scs{\rm L}} - \gamma_{\scs{\rm R}})(J^2 - \gamma_{\scs{\rm L}}\gamma_{\scs{\rm R}})}{2\gamma_{\scs{\rm L}}\gamma_{\scs{\rm R}}J^2} I_{\rm {st}},
\end{equation}
 as follows from Eq.~\eqref{eq:nj_deph}. Hence, we have
\begin{align}
    \begin{split}
    &\bm\alpha^{\dagger} (\hat\rho_{\rm \scs{st}} + \hat{\mathbbm{1}}
    ) \bm\alpha 
    = iB\sum_{{\ell}=1}^{L-1}\bigl(\alpha_{{\ell}+1}^*\alpha_{{\ell}} - \alpha_{{\ell}}^*\alpha_{{\ell}+1}\bigr) \\
    &+(C\!+\!1)|\alpha_1|^2 + (A\!+\!1)\sum_{{\ell}=2}^{L-1}|\alpha_{\ell}|^2 + (D\!+\!1)|\alpha_{\scs{L}}|^2 \;.
   \end{split}
\end{align}
Finally, the Husimi \(Q\)-function is defined as the Fourier transform of the anti‑normally ordered characteristic function~\cite{Gardiner2014}
\begin{equation}\label{Husimi}
    Q({\bm a}, {\bm a^*}) = \frac{1}{\pi^{2L}}\int d^{L}{\bm \alpha}d^{L}{\bm \alpha}^*\:e^{{\bm\alpha}^{\dagger}{\bm a} - {\bm a}^{\dagger}{\bm \alpha}}\chi({\bm \alpha}, {\bm \alpha^*}) \;.
\end{equation}
Before proceeding to current statistics, we emphasize that only the first moments of the Husimi distribution coincide with the expectation values of operators by construction, since the $Q$-representation contains additional vacuum noise. However, statistics derived from Eq.~\eqref{Husimi} is meaningful because it has an appropriate interpretation in the framework of phase-space measurement, such as heterodyne-type detection \cite{Mueller2016}.

\subsection{Current distribution function at \texorpdfstring{$\kappa = 0$}{κ = 0}}}\label{sec:current_dist}
\begin{figure*}
    \centering
    \includegraphics[width=0.95\linewidth, height=0.4\linewidth]{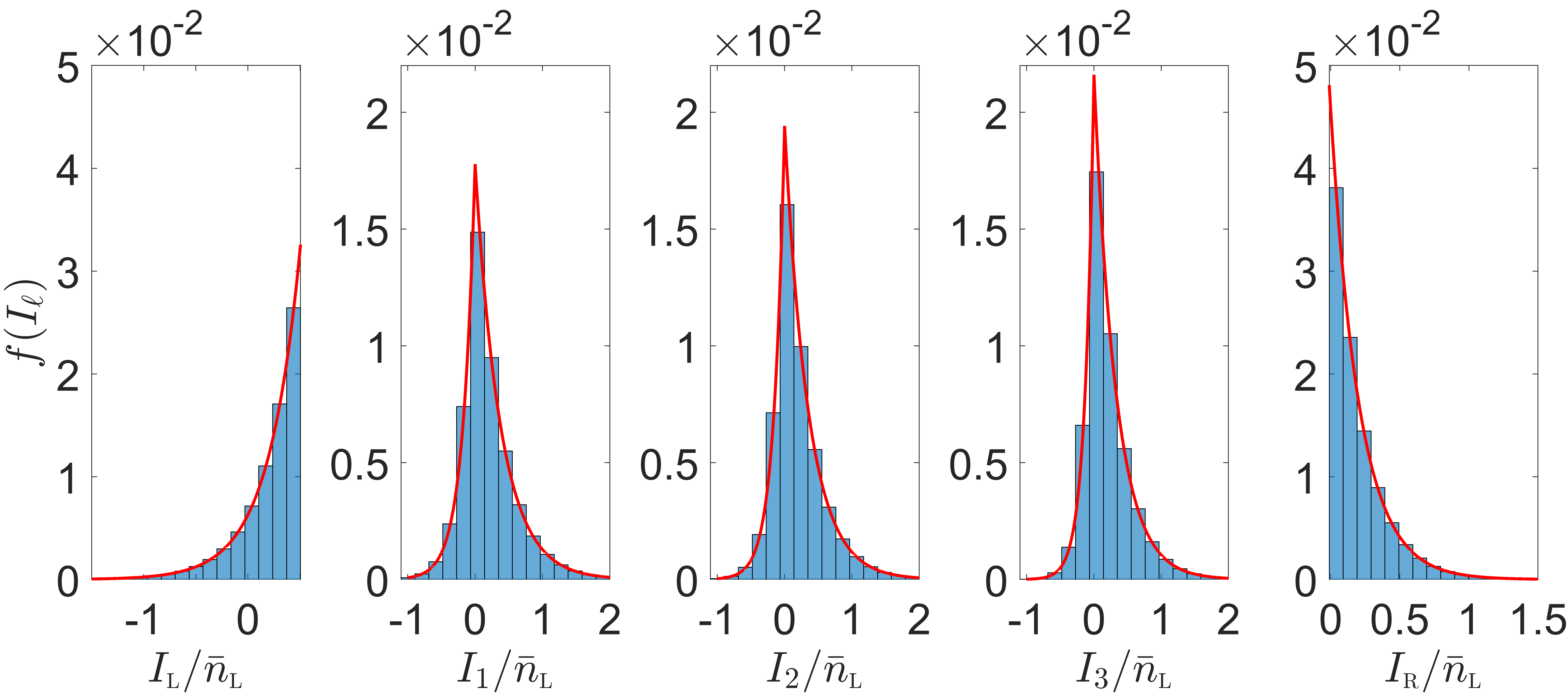}
    \caption{Distribution for the dataset generated from the stochastic differential equations~\eqref{eq:Ito} -- bins, and the analytic distributions~\cref{QI_final,QI_final_R,QI_coherent} -- red solid lines. The parameters are $\bar{n}_{\scs{\rm L}}=100$, $\bar{n}_{\scs{\rm R}}=1$, $\gamma_{\scs{\rm L}}=\gamma_{\scs{\rm R}}=0.5$, $L=4$ and $J=1$. The number of samples is 8800.}
    \label{fig:fig2}
\end{figure*}

Our main object of interest is the stationary current distribution,
\begin{equation}\label{dist_def}
    f(I) = \int d^{L}{\bm a}d^{L}{\bm a}^*\:Q({\bm a}, {\bm a^*})\,\delta\bigl(I-I({\bm a},{\bm a^*})\bigr) \;,
\end{equation}
where \(I({\bm a},{\bm a^*})\) is the pseudo-classical expression for the current, which reads as
\begin{equation}\label{cls_ecur}
    \begin{split}
        I_{\scs{\rm L}} &= \gamma_{\scs{\rm L}}(\bar{n}_{\scs{\rm L}} - a^*_1a_1 + 1), \\
        I_{\scs{\rm R}} &= \gamma_{\scs{\rm R}}(a^*_{\scs{L}}a_{\scs{L}} - \bar{n}_{\scs{\rm R}} - 1 ),
    \end{split}
\end{equation}
for the currents between the reservoirs and the edge sites, and as
\begin{equation}\label{cls_icur}
    I_{\ell} = \frac{J}{2i}\bigl(a_{\ell}^{*} a_{\ell+1} - a_{\ell+1}^{*} a_{\ell}\bigr),
\end{equation}
for the current between sites. The pseudo-classical expressions for current are obtained by using the link between the SPDM elements \eqref{eq:SPDM_def} and the covariant symbols ${\bm a}$ and ${\bm a^*}$ 
\begin{equation}
    \rho_{\ell,\ell'} +\delta_{\ell,\ell'}= \int d^{L}{\bm a}d^{L}{\bm a}^*\:Q({\bm a}, {\bm a^*})\,a^*_{\ell'}a_{\ell} \;.
\end{equation}
Integration with respect to the phase‑space variables can be performed analytically. The details are given in Appendix~\ref{app:QI_deriv}. For the left boundary one obtains the compact result
\begin{equation}\label{QI_final}
\begin{split}
&f(I_{\scs{\rm L}}) = \frac{1}{\gamma_{\scs{\rm L}}(\langle\hat{n}_1\rangle+1)}\,
\exp\!\left(\frac{I_{\scs{\rm L}} - \gamma_{\scs{\rm L}}(\bar{n}_{\scs{\rm L}}+1)}
{\gamma_{\scs{\rm L}}(\langle\hat{n}_1\rangle+1)}\right)\,\times \\
&\Theta\bigl(\gamma_{\scs{\rm L}}(\bar{n}_{\scs{\rm L}}+1) - I_{\scs{\rm L}}\bigr).
\end{split}
\end{equation}
This is an exponential distribution with a sharp cutoff at the maximal possible current \(I_{\max} = \gamma_{\scs{\rm L}}(\bar{n}_{\scs{\rm L}}+1)\).
The origin of the cutoff can be understood from Eq.~\eqref{cur_res}, which states that the current never exceeds $\gamma_{\scs{\rm L}}\bar{n}_{\scs{\rm L}}$.
However, since we work in the Husimi function formalism, the cut-off in the derived formulas is shifted by $\gamma_{\scs{\rm L}}$ due to vacuum noise.
A completely analogous calculation for the right boundary yields
\begin{equation}\label{QI_final_R}
\begin{split}
&f(I_{\scs{\rm R}}) = \frac{1}{\gamma_{\scs{\rm R}}(\langle\hat{n}_{\scs{L}}\rangle+1)}\,
\exp\!\left(-\frac{I_{\scs{\rm R}} + \gamma_{\scs{\rm R}}(\bar{n}_{\scs{\rm R}}+1)}
{\gamma_{\scs{\rm R}}(\langle\hat{n}_{\scs{L}}\rangle+1)}\right)\,\times \\
&\Theta\bigl(\gamma_{\scs{\rm R}}(\bar{n}_{\scs{\rm R}}+1) + I_{\scs{\rm R}}\bigr).
\end{split}
\end{equation}
Finally, the distribution for current between two neighboring inner sites can be obtained by a similar procedure, see Appendix~\ref{app:QII_deriv}. The final result takes the following form
\begin{equation}\label{QI_coherent}
f(I_{\ell}) = \frac{1}{Ju_\ell}\left\{
\begin{array}{ll}
\exp\left(-\dfrac{2I_{\ell}}{Ju_\ell + I_{\rm st}}\right), & I_{\ell} \ge 0, \\[8pt]
\exp\left(\dfrac{2I_{\ell}}{Ju_\ell - I_{\rm st}}\right), & I_{\ell} \le 0,
\end{array} 
\right.
\end{equation}
where 
\begin{equation}
u_\ell=\sqrt{(\langle\hat{n}_{\ell}\rangle+1)(\langle\hat{n}_{\ell+1}\rangle+1)}.
\end{equation}
In Eqs.~\eqref{QI_final}--\eqref{QI_coherent} quantities $\langle\hat{n}_{1}\rangle$, $\langle\hat{n}_{\ell}\rangle$, $\langle\hat{n}_{\scs{L}}\rangle$ and $I_{\rm st}$ are given by Eqs.~\eqref{eq:I_deph}--\eqref{eq:nj_deph} in Sec.~\ref{sec:II}.
For the  inter-site current, the result~\eqref{QI_coherent} is an asymmetric Laplace distribution. The asymmetry is entirely governed by the mean current $I_{\rm st}$. The denominators contain \(Ju_\ell \pm I_{\rm st}\), and when \(I_{\rm st} \to 0\) the distribution becomes symmetric,
\begin{equation}
f(I) = \frac{1}{Ju_{{\ell} }}\exp\left(-2\frac{|I|}{Ju_{\ell}}\right).
\end{equation}
For \(I_{\rm st} \neq 0\) the asymmetry is present for any bond, regardless of whether their populations are equal. 

To exemplify the analytic predictions, we generated a data-set by solving the stochastic differential equations that correspond to the Husimi $Q$-function. The approach is based on the equivalence between the Fokker–Planck equation, which is the Fourier transform of Eq.~\eqref{eq:chi_eq} for $\kappa=0$, and the following set of It\^{o} drift-diffusion equations for covariant symbols $a_\ell(t)$
\begin{align}
    %\begin{split}
        da_1 &= \frac{1}{2}\Bigl(iJ a_2 - \gamma_{\scs{\rm L}}a_1\Bigr)dt + \sqrt{\gamma_{\scs{\rm L}}(\bar n_{\scs{\rm L}}+1)}\;d\xi_{\scs{\rm L}}, \nonumber\\[8pt]
        da_\ell &= \frac{iJ}{2}\bigl(a_{\ell+1}+a_{\ell-1}\bigr)dt, \quad \ell=2,\dots,L-1,\label{eq:Ito} \\[8pt]
        da_{\scs{L}} &= \frac{1}{2}\Bigl(iJ a_{\scs{L}-1} - \gamma_{\scs{\rm R}}a_{\scs{L}}\Bigr)dt + \sqrt{\gamma_{\scs{\rm R}}(\bar n_{\scs{\rm R}}+1)}\;d\xi_{\scs{\rm R}}.\nonumber\
    %\end{split}
\end{align}
where $d\xi_{\ell}$ is complex white noise that satisfies $\langle d\xi_{\ell}\rangle=0$, $\langle d\xi_{\ell} d\xi_{\ell'}\rangle = 0$, and $\langle d\xi_{\ell}^* d\xi_{\ell'}\rangle = \delta_{\ell,\ell'} dt$. The deterministic evolution of the inner sites in Eq.~\eqref{eq:Ito} reflects the absence of both dephasing and direct coupling to the reservoirs. In simulations, we numerically trace many trajectories in parallel. After exceeding the relaxation time $T_{\rm r}$, we record $a_\ell(T_{\rm r})$ and compute the instantaneous current with the pseudo-classical formulas appropriate for the bond of interest, Eq.~\eqref{cls_ecur} and Eq.~\eqref{cls_icur}. The resulting dataset is built on $N=8800$ independent realizations. The estimate for the relaxation time is  obtained from analysis of the spectrum of the effective non-Hermitian Hamiltonian as the life-time of the most long-lived resonance. For the case of equal relaxation constants, $\gamma_{\scs{\rm L}} = \gamma_{\scs{\rm R}}$, it can be found that $T_r \propto L^3$, see Appendix~\ref{app:QIII_spec}, which consistent with the earlier findings in \cite{ Guo2017, Znidaric2015}. The results are visualized in Fig.~\ref{fig:fig2}.

\subsection{Current-current correlation function at \texorpdfstring{$\kappa = 0$}{κ = 0}}
The Gaussian nature of the steady-state for $\kappa=0$ not only allows us to compute the current distribution but also the correlation between currents at different bonds. Such correlations characterize the spatial structure of fluctuations. The product $\hat {\cal I}_\ell \hat {\cal I}_{\ell'}$ is a fourth-order polynomial in field operators. Applying~\eqref{eq:moments} after anti-normal ordering, which is consistent with the definition of the characteristic function, we obtain
\begin{equation}
\begin{split}
        \langle \hat{\cal I}_{\ell}\hat{\cal I}_{\ell'} \rangle &= \frac{J^2}{4}\left[ 4B^2 \!-\! M_{{\ell'}\!+1,\ell}M_{\ell+1,{\ell'}}\right.  \\
        &\!+\! M_{{\ell'},\ell}M_{\ell+1,{\ell'}+1} \!+\! M_{{\ell'}+1,\ell+1}M_{\ell,{\ell'}} \\
        &\!-\! \left.M_{{\ell'},\ell+1}M_{\ell,{\ell'}+1} \!+\! \delta_{\ell,{\ell'}}(M_{\ell,\ell} \!+\! M_{\ell+1,\ell+1})\right],
\end{split}
\end{equation}
where $M_{\ell,{\ell'}}$ are the elements of $\hat{M}$, and $B=I_{\rm st}/J$. Using $M_{\ell,\ell} = (\langle n_\ell\rangle+1)$, we obtain the compact form
\begin{equation}
    \begin{split}
        \langle \hat{\cal I}_\ell \hat{\cal I}_{\ell'} \rangle =
        \begin{cases}
        \dfrac{3}{2}I_{\rm st}^2+\dfrac{J^2}{4}v_\ell^2, & \ell = \ell', \\[8pt]
        \dfrac{3}{2}I_{\rm st}^2, & |\ell - \ell'| = 1, \\[8pt]
        I_{\rm st}^2, & |\ell - \ell'| \ge 2,
        \end{cases}
    \end{split}
\end{equation}
where
\begin{equation}
v^2_\ell=\langle\hat{n}_{\ell}\rangle + 
        \langle\hat{n}_{\ell + 1}\rangle +
2\langle\hat{n}_{\ell}\rangle\langle\hat{n}_{\ell+1}\rangle.
\end{equation}
It should also be mentioned that the current-current correlation function $\langle \hat{\cal I}_\ell \hat{\cal I}_{\ell'} \rangle$ can be obtained directly from the master equation~\eqref{m_eq}, since the state is Gaussian and all four-point correlation functions factorize according to Wick's theorem. Several observations are in order.  
(i) For $|\ell-\ell'|=1$ the covariance ${\rm cov}(\hat{\cal I}_{\ell},\hat{\cal I}_{\ell+1})=I_{\rm st}^2/2$. This positive correlation between neighboring currents is due to the shared site, through which fluctuations of the local occupation influence both bonds coherently. 
(ii) For $|\ell-\ell'|\ge2$ the correlation reduces to the product of the mean currents, $\langle\hat{\cal I}_{\ell}\hat{\cal I}_{\ell'}\rangle=I_{\rm st}^2$. The covariance is zero, i.e. the fluctuations at sites separated by more than one bond are completely uncorrelated. (iii) In the limit of zero bias, the inter-bond correlations vanish, while the autocorrelation remains finite and is given by the thermal noise.

One may also consider the correlation between the boundary current and the inner bond current. At the left edge, we obtain
\begin{equation}
     \langle \hat{\cal I}_{\scs{\rm L}} \hat{\cal I}_{\scs{\rm L}} \rangle =I_{\rm st}^2 + \gamma^2_{\scs{\rm L}}\Big(\langle \hat{n}_1\rangle^2 + \langle \hat{n}_1\rangle\Big), 
\end{equation}
while for the correlators for the first inner bond, $\ell=1$, and for more distant inner bonds, $\ell \ge 2$,
\begin{equation}
    \begin{split}
        \langle \hat{\cal I}_{\scs{\rm L}} \hat{\cal I}_{\ell} \rangle =
        \begin{cases}
        I_{\rm st}^2 - I_{\rm st}\gamma_{\scs{\rm L}}\Big(\langle \hat{n}_1\rangle + \dfrac{1}{2}\Big), & \ell  = 1, \\[8pt]
        I_{\rm st}^2, & \ell \ge 2.
        \end{cases}
    \end{split}
\end{equation}
Similarly, at the right edge,
\begin{equation}
     \langle \hat{\cal I}_{\scs{\rm R}} \hat{\cal I}_{\scs{\rm R}} \rangle = I_{\rm st}^2 + \gamma^2_{\scs{\rm R}}\Big(\langle \hat{n}_{\scs{L}}\rangle^2 + \langle \hat{n}_{\scs{L}}\rangle\Big),
\end{equation}
and the correlation with the inner bond current is given by
\begin{equation}
    \begin{split}
        \langle \hat{\cal I}_{\ell} \hat{\cal I}_{\scs{\rm R}} \rangle=
        \begin{cases}
        I_{\rm st}^2 + I_{\rm st}\gamma_{\scs{\rm R}}\Big(\langle \hat{n}_{\scs{L}}\rangle + \dfrac{1}{2}\Big), &\ell  = L-1, \\[8pt]
        I_{\rm st}^2, &\ell \le L\!-\!2.
        \end{cases}
    \end{split}
\end{equation}
The difference in sign between the left and right boundaries reflects our definition of the current direction, see~Fig.~\ref{fig:scheme}. 

\subsection{Current statistics at \texorpdfstring{$\kappa \neq 0$}{κ neq 0}}
\begin{figure*}[ht]
    \centering
    \includegraphics[width=0.95\linewidth]{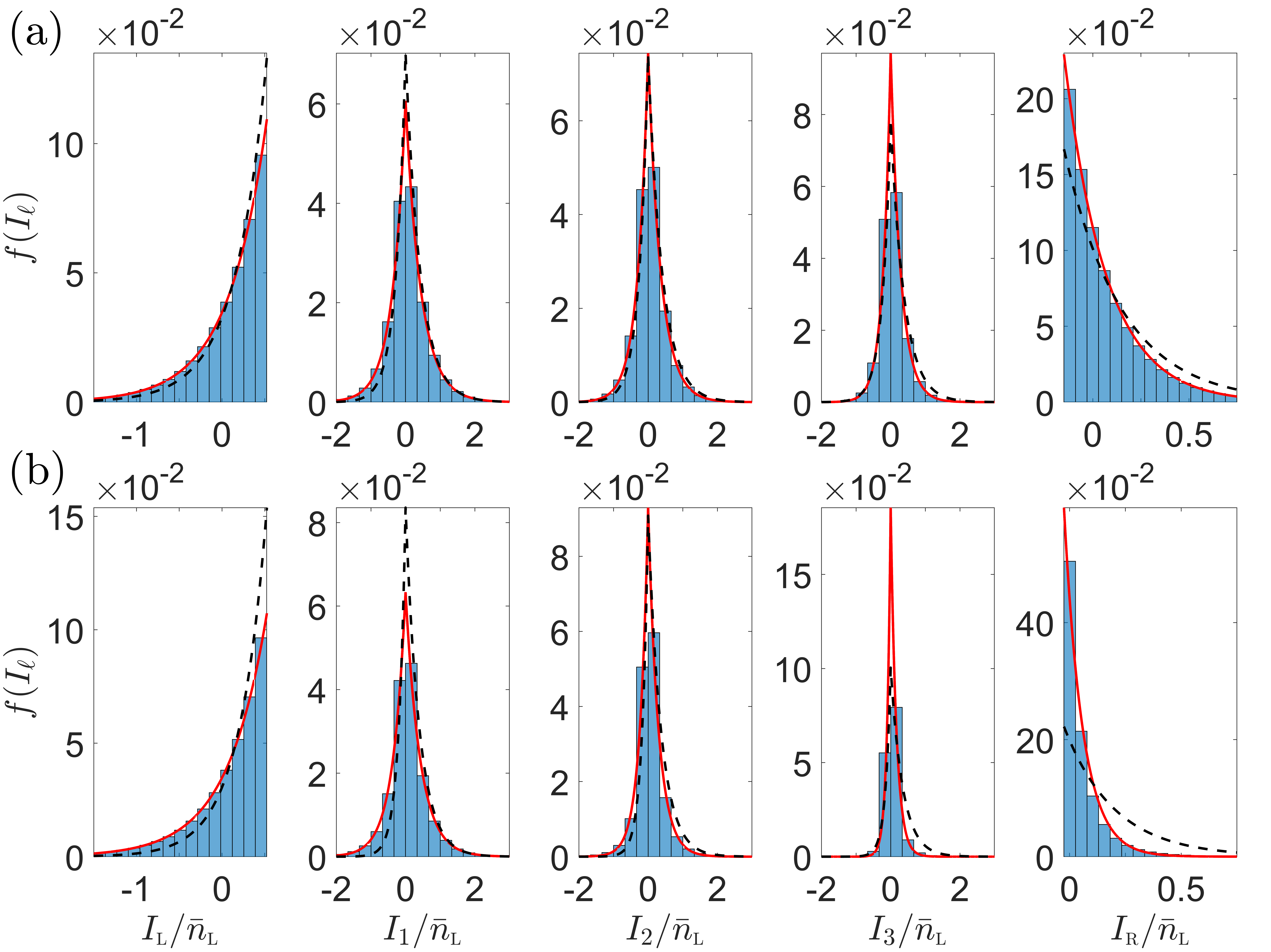}
    \caption{Distribution for the dataset generated from the stochastic differential equations~\eqref{eq:SDE_with_dephasing} -- bins, and the analytic distributions with $\kappa=0$~\cref{QI_final,QI_final_R,QI_coherent} -- black dashed lines, and same distributions but with non-zero dephasing -- red solid lines. The parameters are $\bar{n}_{\scs{\rm L}}=20$, $\bar{n}_{\scs{\rm R}}=5$, $\gamma_{\scs{\rm L}}=\gamma_{\scs{\rm R}}=0.5$, $L=4$, $\kappa = 1$ and $J=1$ row (a); $\bar{n}_{\scs{\rm L}}=20$, $\bar{n}_{\scs{\rm R}}=0$, $\gamma_{\scs{\rm L}}=\gamma_{\scs{\rm R}}=0.5$, $L=4$, $\kappa = 2$ and $J=1$ row (b). The number of samples is 8800.}
    \label{fig:fig3}
\end{figure*}
In the case of non-zero phase damping, the Gaussian Anzatz for the characteristic function does not satisfy Eq.~\eqref{eq:chi_eq}. Nevertheless, the SPDM presented in Sec.~\ref{sec:II} is an exact solution. The SPDM is positive and, therefore, applying it in Eq.~\eqref{chi_gauss} and  Eq.~\eqref{ch_fun_sol} results in a positive Husimi distribution, which can be used for deriving the distributions of current. The derivation results in the same expressions, Eqs.~\eqref{QI_final}--\eqref{QI_coherent}, with the only difference that Eqs.~\eqref{eq:I_deph}--\eqref{eq:nj_deph} are now evaluated with non-zero $\kappa$. To access the accuracy of this {\it Gaussian approximation} we applied the modified equations of motion for the covariant symbols, that account for phase damping via multiplicative noise,
\begin{equation}
\begin{aligned}
da_\ell &= \frac{1}{2}\biggl[iJ(a_{\ell-1}+a_{\ell+1}) - (\gamma_{\ell}+\kappa)a_\ell\biggr]dt \\
&\quad + \sqrt{\gamma_\ell(\bar n_\ell+1)}\,d\xi_\ell + i\sqrt{\kappa_\ell}\,a_\ell\,d\eta_\ell,
\end{aligned}
\label{eq:SDE_with_dephasing}
\end{equation}
where $\ell = 1,\dots,L$, $a_0 = a_{\scs{L}+1} = 0$, $\gamma_\ell = \bar n_\ell = 0$ for bulk sites, and $d\eta_\ell$ are independent Wiener processes with $\langle d\eta_\ell d\eta_k\rangle = \delta_{\ell k}dt$.

To assess the validity of analytic current distributions beyond the quasi-free limit, we performed numerical simulations with the stochastic differential equations~\eqref{eq:SDE_with_dephasing} for non-‑zero dephasing. Although the steady state is no longer Gaussian, and the Gaussian characteristic function~\eqref{chi_gauss} does not solve the corresponding equation, we find that the analytic formulas for {the current distributions} derived for the Gaussian steady state~\cref{QI_final,QI_final_R,QI_coherent} remain remarkably accurate when supplied with the exact steady‑state values of the second moments obtained for the problem with dephasing. In Fig.~\ref{fig:fig3} we compare the histograms from the stochastic simulations  with the original zero‑dephasing analytic distributions as well as with the analytic distributions evaluated in the Gaussian approximation using the exact SPDM for non‑zero dephasing. The numerical data are in good agreement with the latter for both boundary and inter‑site currents.
%This agreement holds for a range of parameters and chain lengths up to 
%$L=10$; longer chains were not tested numerically.

\section{Conclusion}
We have considered quantum transport in a tight‑binding bosonic chain coupled to Markovian reservoirs at its boundaries and subject to uniform on‑site dephasing. 
%Using the master‑equation approach, we derived a closed set of equations for the single‑particle density matrix, which allowed us to obtain exact analytic expressions for the stationary current and the site populations for arbitrary dephasing strength and boundary driving. The expression for the current given by Eq.~\eqref{eq:I_deph} constitutes the central result of the paper. It is noteworthy that the obtained solution describes transition from the ballistic, $\kappa=0$, to diffusive, $\kappa\neq 0$, transport. Although presented for transport of bosons, this solution also directly applies to non-interacting fermionic chains, similar to that considered in \cite{lacerda2021}, see discussion in Appendix ~\ref{app:D}.
In the quasi‑free limit of vanishing dephasing, the stationary state is Gaussian.
Starting from the analytic expression for the single‑particle density matrix \cite{Asadian2013}, the phase‑space formalism immediately provides the explicit characteristic function of the steady-state.  This function grants access to correlation functions of any order obtained via heterodyne-type detection.  
In particular, it allowed us to derive the probability distributions of the boundary and inter‑site currents.  
The boundary current follows an exponential distribution with a sharp cutoff, while the coherent current between neighboring sites exhibits an asymmetric Laplace distribution. It is worth mentioning that exponential distributions of probability current components were previously reported for quantum transport through chaotic billiards \cite{saichev2002statistics}.
The Gaussian nature of the steady-state also gave access to the current–current correlation functions. 
We showed that the current fluctuations on distant bonds are uncorrelated, whereas the neighboring bonds exhibit a positive covariance. For the case of non-zero dephasing, we constructed a Guassian approximation, which relies on the exact analytic solution for the single particle density matrix. It is demonstrated numerically that the Guassian approximation describes current distributions to a good accuracy.  

As pathways for engineering Markovian baths with ultracold bosonic gases are recently suggested in \cite{Wu2022a}, we believe that our work provides a comprehensive characterization of current fluctuations in open quantum systems connected to such baths. 
Natural extensions include the investigation of full counting statistics in the presence of dephasing, the analysis of higher‑order cumulants, and the exploration of similar distributions in interacting fermionic or bosonic chains.

{\em Acknowledgments.} 
This work was supported by the Russian Science Foundation under Grant No. 25-12-00268.

\bibliography{mybib}

\appendix
\section{Current distribution for the external bonds}\label{app:QI_deriv}
Here we derive the distribution for the current between the left reservoir and the first site in the chain. We start by rewriting Eq.~\eqref{dist_def}
\begin{equation}\label{ }
    f(I) = \int d^{L}{\bm a}d^{L}{\bm a}^*\:Q({\bm a}, {\bm a^*})\,\delta\bigl(I-I({\bm a},{\bm a^*})\bigr) \;,
\end{equation}
which is the starting point in the derivation. For brevity, we shall use $I_{\rm L}=I$ in this section. In what follows, the limits of all integrals are dropped  and integration is performed from minus to plus infinity.
In the first step, we apply the Fourier representation of the delta-function
\begin{equation}
    \delta\bigl(I-I({\bm a},{\bm a^*})\bigr) = \frac{1}{2\pi}\int d\omega\, e^{i\omega(I-I({\bm a},{\bm a^*}))} \;,
\end{equation}
and the definition of the Husimi function \eqref{Husimi} to obtain
\begin{align}
\begin{split}\label{QI_sub}
f(I) &= \frac{1}{2\pi}\int d\omega\, e^{i\omega I} 
\frac{1}{\pi^{2L}} \int d^{L}{\bm \alpha}d^{L}{\bm \alpha}^*d^{L}{\bm a}d^{L}{\bm a}^* 
\\[12pt]
&\qquad  
\times\chi(\bm{\alpha},\bm{\alpha}^*) e^{{\bm\alpha}^{\dagger}{\bm a} - {\bm a}^{\dagger}{\bm \alpha}} 
e^{-i\omega I({\bm a},{\bm a^*})}.
\end{split}
\end{align}
The pseudo-classical formula for the current is given by Eq.~\eqref{cls_ecur}
\begin{equation}\label{Icl_def}
I({\bm a},{\bm a^*}) =  \gamma_{\scs{\rm L}}(\bar{n}_{\scs{\rm L}} - a^*_1a_1 + 1).
\end{equation}
We can now evaluate the integrals by changing the order of integration  starting with the \(\bm{a}\)-integrals.
For any site with \(\ell \neq 1\) we have
\begin{equation}
\int d^2 a_{\ell}\, e^{\alpha_{\ell}^* a_{\ell} - a_{\ell}^*\alpha_{\ell}} 
= \pi^2 \delta^{(2)}(\alpha_{\ell}),
\end{equation}
where \(\delta^{(2)}(\alpha_{\ell})=\delta(\mathrm{Re}\,\alpha_{\ell})\delta(\mathrm{Im}\,\alpha_{\ell})\).
Integrating with respect to \(a_2,\dots,a_{\scs{L}}\) and subsequently with respect to \(\alpha_2,\dots,\alpha_{\scs{L}}\) simply yields a factor \(\pi^{2(L-1)}\).
Because off‐diagonal terms that mix \(\alpha_{\ell},\alpha_{\ell'}^*\) for $\ell\neq\ell'$ are forced to vanish, the characteristic function reduces to
\begin{equation}
\chi_{\rm red}(\alpha_1,\alpha_1^*) = \exp\bigl(-g|\alpha_1|^2\bigr),
\end{equation}
with \(g = \langle\hat{n}_1\rangle + 1\). The remaining integral with respect to \(a_1,a_1^*\) includes the phase factor emerging from the formula for current
\begin{equation}
\tilde{I}(\alpha_1,\omega) = \int d a_1 d a_1^*\, 
e^{\alpha_1^* a_1 - a_1^*\alpha_1 + i\omega\gamma_{\scs{\rm L}} |a_1|^2}.
\end{equation}
Switching to real quadratures
\begin{equation}
a_1 = (q+ip)/\sqrt{2}, \ \ d a_1 d a_1^* = \frac12 dq\,dp,
\end{equation}
we obtain a product of two Fresnel integrals leading to
\begin{equation} \label{dfdf}
\tilde{I}(\alpha_1,\omega) = \frac{\pi i}{\omega\gamma_{\scs{\rm L}}}\, 
\exp\!\left(-\frac{i}{\omega\gamma_{\scs{\rm L}}}|\alpha_1|^2\right).
\end{equation}
Using Eq.~\eqref{dfdf} in Eq.~\eqref{QI_sub} gives
\begin{equation}
\begin{split}
f(I) = \frac{1}{2\pi}\int d\omega\, 
e^{i\omega[I - \gamma_{\scs{\rm L}}(\bar{n}_{\scs{\rm L}}+1)]}\, \\
\times\frac{ i}{\pi\omega\gamma_{\scs{\rm L}}}\int d\alpha_1d\alpha^*_1\, e^{-\left[g + i/(\omega\gamma_{\scs{\rm L}})\right]|\alpha_1|^2}.
\end{split}
\end{equation}
The Gaussian integral with respect to \(\alpha_1,\alpha^*_1\) is evaluated as 
\begin{equation}
\int d\alpha_1 d\alpha_1^*\, e^{-\left[g + i/(\omega\gamma_{\scs{\rm L}})\right]|\alpha_1|^2}
= \frac{\pi}{g + i/(\omega\gamma_{\scs{\rm L}})},
\end{equation}
which leads to
\begin{equation}
f(I) = \frac{i}{2\pi\gamma_{\scs{\rm L}}g}\int d\omega\, 
\frac{e^{i\omega[I - \gamma_{\scs{\rm L}}(\bar{n}_{\scs{\rm L}}+1)]}\,}{\omega + i/(\gamma_{\scs{\rm L}}g)}.
\end{equation}
Finally, evaluating the above integral by the residue theorem and recalling that \(g = \langle\hat{n}_1\rangle + 1\), we arrive at the first line of Eq.~\eqref{QI_final} of the main text
\begin{equation}
\begin{split}
&f(I) = \frac{1}{\gamma_{\scs{\rm L}}(\langle\hat{n}_1\rangle+1)}\,
\exp\!\left(\frac{I - \gamma_{\scs{\rm L}}(\bar{n}_{\scs{\rm L}}+1)}
{\gamma_{\scs{\rm L}}(\langle\hat{n}_1\rangle+1)}\right)\, \\
&\times\Theta\bigl(\gamma_{\scs{\rm L}}(\bar{n}_{\scs{\rm L}}+1) - I\bigr).
\end{split}
\end{equation}
The derivation for the bond connected to the right reservoir is essentially the same, and, therefore, omitted for brevity.

\section{Current distribution for the internal bonds}\label{app:QII_deriv}

Now we derive the probability distribution of the stationary coherent current between two adjacent sites \(\ell\) and \(\ell+1\).  
The pseudo-classical formula for the current, Eq.~\eqref{cls_icur} is as follows
\begin{equation}\label{Icl_coherent2}
I(\bm{a}) = \frac{J}{2i}\bigl(a_{\ell}^{*} a_{\ell+1} - a_{\ell+1}^{*} a_{\ell}\bigr).
\end{equation}
The initial step is the same as in Appendix A.
Because the current depends only on the covariant symbols at the \(\ell\) and \(\ell+1\) sites, we integrate with respect to all other variables.  
The marginal \(Q\)-function for the two relevant sites is a Gaussian distribution. This can be shown by performing the integrals with respect to \(\alpha_{\ell'}\) and \(a_{\ell'}\) for \(\ell'\neq \ell,\ell+1\) exactly as in Appendix A. The delta functions fix \(\alpha_\ell=0\) and the remaining integrations yield a two‑mode Gaussian
\begin{equation}\label{Q2}
\begin{split}
    &Q_{\scs{\rm red}}(a_\ell,a_{\ell'+1},a^*_\ell,a^*_{\ell'+1}) = \\
    &\frac{1}{\pi^{2}\det\hat\Sigma}\,
    \exp\!\left[-\bigl(a_\ell^{*},\,a_{\ell+1}^{*}\bigr)\,\hat\Sigma^{-1}
\begin{pmatrix} a_\ell \\[2pt] a_{\ell+1} \end{pmatrix}\right],
\end{split}
\end{equation}
where the matrix \(\hat\Sigma\) is the restriction of \(\hat{\rho}_{\scs{\rm st}}+\mathbbm{1}\) to the two relevant sites.  
With the explicit form \eqref{eq:SPDM_form} we have
\begin{equation}
    \hat\Sigma = 
    \begin{pmatrix}
        \langle\hat{n}_{\ell}\rangle+1 & -iB \\[4pt]
        iB & \langle\hat{n}_{\ell+1}\rangle+1
    \end{pmatrix}
    .
\end{equation}
The current distribution can now be written as
\begin{align}
\begin{split}
    f(I) = \int da_\ell\,da_{\ell+1}da^*_\ell\,da^*_{\ell+1} \\  \times Q_{\rm red}(a_\ell,a_{\ell+1},a^*_\ell,a^*_{\ell+1})\;
\delta\bigl(I - I(\bm{a})\bigr).
\end{split}
\end{align}
Using the Fourier integral representation of the \(\delta\)-function,
\begin{equation}
\delta\bigl(I - I(\bm{a})\bigr) = \frac{1}{2\pi}\int d\omega\,
e^{i\omega(I - I(\bm{a}))},
\end{equation}
and exchanging the order of integration, we obtain
\begin{equation}\label{QI_coherent_FT}
f(I) = \frac{1}{2\pi}\int d\omega\, e^{i\omega I}\,
\tilde{\chi}(\omega),
\end{equation}
with the characteristic function of the current
\begin{align}\label{chired}
\begin{split}
\tilde{\chi}(\omega) = \int da_\ell\,da_{\ell+1}da^*_\ell\,da^*_{\ell+1}\; \\
\times Q_{\rm red}(a_\ell,a_{\ell+1})\, e^{-i\omega I(\bm{a})}.
\end{split}
\end{align}
Evaluating the above integral with the use of Eq.~\eqref{Icl_coherent2} we obtain
\begin{equation}
\tilde{\chi}(\omega) = \frac{4}{J^{2}\det\hat\Sigma }\,
\cdot\frac{1}{(\omega - \omega_1)(\omega - \omega_2)},
\end{equation}
where 
\begin{align}
    \begin{split}
    &\omega_{1,2} = i\lambda_{1,2},\\
    &\lambda_1 = \frac{2(u_{\ell}-B)}{J\det\hat\Sigma } > 0, \\
    &\lambda_2 = -\frac{2(u_{\ell}+B)}{J\det\hat\Sigma } < 0, \\
    &u_{\ell} =  \sqrt{(\langle\hat{n}_\ell\rangle+1)(\langle\hat{n}_{\ell+1}\rangle+1)}.
    \end{split}
\end{align}
Now, the Fourier integral \eqref{QI_coherent_FT} is evaluated by the residue theorem closing the contour in the complex \(\omega\)-plane. The result reads
\begin{equation*}%\label{QI_coherent_final}
f(I) = \frac{1}{Ju_\ell}\,
\begin{cases}
\exp\!\left(-\dfrac{2I}{Ju_\ell + I_{\rm st}}\right), & I \ge 0,\\[12pt]
\exp\!\left(\dfrac{2I}{Ju_\ell - I_{\rm st}}\right), & I \le 0,
\end{cases}
\end{equation*}
which is Eq.~\eqref{QI_coherent} from the main text.

\section{Spectrum of the effective non-Hermitian Hamiltonian}\label{app:QIII_spec}
We only consider the simplest case of Eq.~\eqref{eq:Heff} when the dephasing is absent $\kappa=0$ and the left and right relaxation rates are equal to each other $\gamma_{\rm L}=\gamma_{\rm R}:=\gamma$, which makes the problem symmetric. We apply a perturbative approach to find the solution of the eigenvalue problem for arbitrary $L$
\begin{equation}\label{Apa1}
  (\widehat H_{\rm eff}-E_p)|\psi_p\rangle = 0,
\end{equation}
in the form of a power series in ${\gamma}/{J}$. Let us start by writing Eq.~(\ref{Apa1}) in the matrix form. The effective Hamiltonian reads
\begin{equation}\label{Apa2}
\widehat H_{\rm eff}= -\frac{1}{2}
\begin{pmatrix}
i\gamma & J & & & 0 \\
J & 0 & J & & \\
& \ddots & \ddots & \ddots & \\
& & J & 0 & J \\
0 & & & J & i\gamma
\end{pmatrix}
\end{equation}
so that the eigenvalue equation becomes
\begin{equation}
  -\frac{J}{2}\bigl(\psi_{\ell-1}+\psi_{\ell+1}\bigr) = E\,\psi_\ell,\qquad \ell=2,\ldots,L-1,
\end{equation}
with the boundary conditions
\begin{align}
 & -i\frac{\gamma}{2}\,\psi_1-\frac{J}{2}\,\psi_2 = E\,\psi_1,\label{Apa2b}\\
 & -i\frac{\gamma}{2}\,\psi_L-\frac{J}{2}\,\psi_{L-1} = E\,\psi_L,\label{Apa2c}
\end{align}
where $\psi_\ell=\langle \ell| \psi \rangle$.
\begin{figure}
    \centering
    \includegraphics[width=1.0\linewidth]{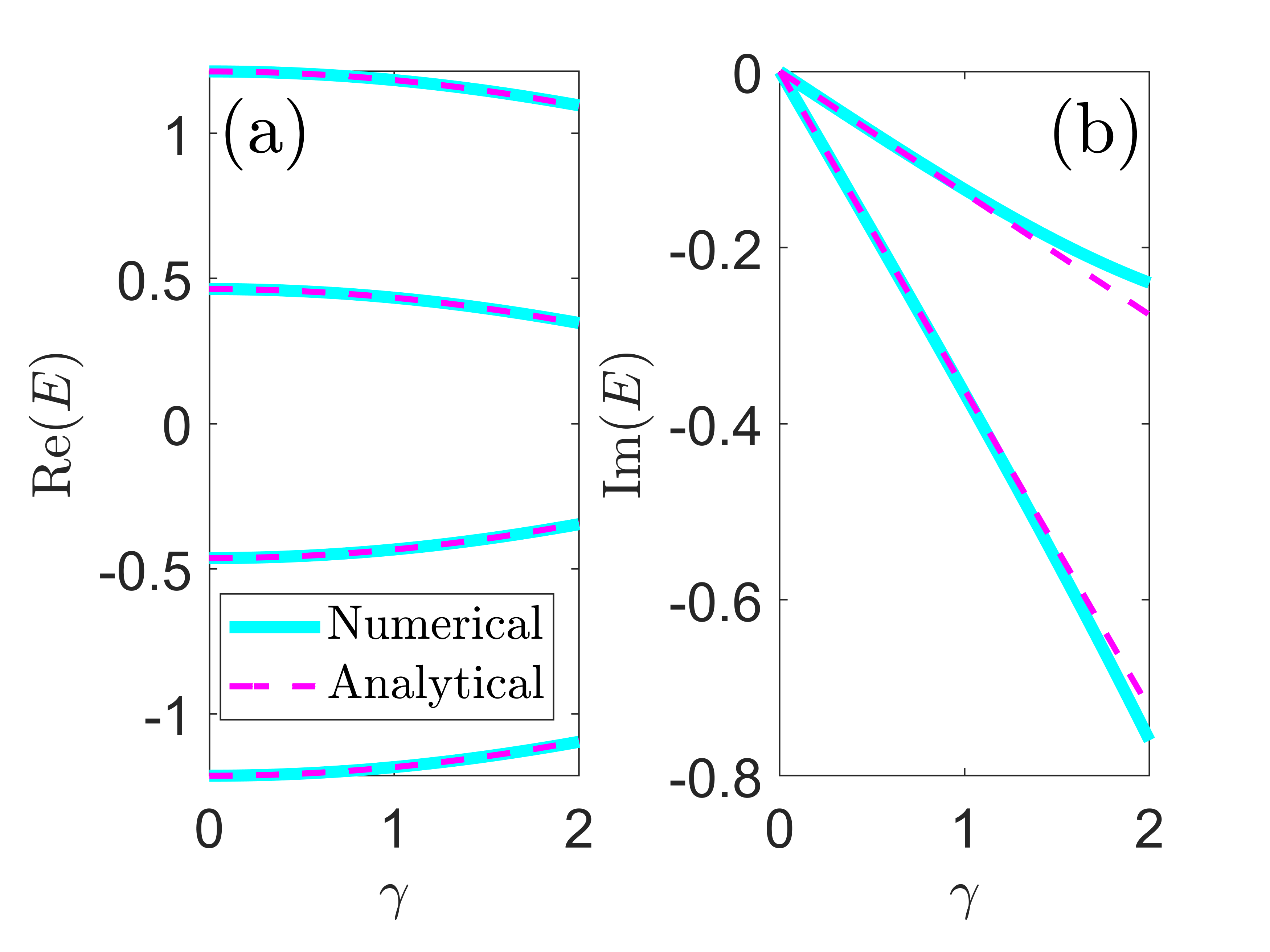}
    \caption{Real (a) and imaginary (b) parts of the energy spectrum of the effective non‑Hermitian Hamiltonian for \(L=4\), \(J=1.5\) against the dissipation strength \(\gamma\). Dashed lines represent the perturbative result up to \(\mathcal O^2(\gamma/J)\), see Eq.~(\ref{Energy}). The solid lines show the eigenvalues obtained from exact numerical diagonalization.}
    \label{fig:sup_3}
\end{figure}
Let us find a solution in the form of a superposition of two counter‑propagating waves
\begin{equation}\label{Apa3}
\psi_\ell = Ae^{ik\ell} + Be^{-ik\ell},
\end{equation}
with the dispersion relation
\begin{equation}\label{Apa4}
E_{} = -J\cos(k).
\end{equation}
It is easy to see that Eq.~(\ref{Apa3}) solves the bulk equation. For Eq.~(\ref{Apa3}) to be an eigenmode, the boundary conditions (\ref{Apa2b})--(\ref{Apa2c}) must also be satisfied. This yields a homogeneous linear system for the coefficients $A$ and $B$. Requiring a non‑trivial solution gives the quantization condition
\begin{equation}\label{Apa8}
\sin[k(L+1)] - 2i\frac{\gamma}{J}\sin(kL) - \frac{\gamma^2}{J^2}\sin[k(L-1)] = 0.
\end{equation}
Now we introduce a series expansion of $k$ in powers of ${\gamma}/{J}$,
\begin{equation}\label{Apa9}
k = k^{(0)} + \alpha \frac{\gamma}{J} + \beta \frac{\gamma^2}{J^2} + \mathcal{O}^3\left(\frac{\gamma}{J}\right),
\end{equation}
where the zeroth‑order wave numbers are those of the closed chain,
\begin{equation}\label{Apa10}
k^{(0)}_{p} = \frac{\pi p}{L+1}, \qquad p = 1,2,\ldots,L.
\end{equation}
Substituting Eq.~(\ref{Apa9}) into Eq.~(\ref{Apa8}) and collecting terms of the same order in ${\gamma}/{J}$, one finds
\begin{align}
\alpha_{p} &= -\frac{2i}{L+1}\sin\left(\frac{\pi p}{L+1}\right),\label{alpha}\\
\beta_{p}  &= \frac{L-1}{(L+1)^2}\sin{\left(\frac{2\pi p}{L+1}\right)}. \label{beta}
\end{align}
With the dispersion relation (\ref{Apa4}) we obtain the energy spectrum up to second order in $\gamma$
\begin{align}\label{Energy}
\begin{split}
        E_{p} = &-J\cos k^{(0)}_{p} - i\frac{2\gamma}{L+1}\sin^2 k^{(0)}_{p} \\
        &+ \frac{2(L-2)\gamma^2}{J(L+1)^2}\sin^2 k^{(0)}_{p} \cos k^{(0)}_{p} + \mathcal{O}(\gamma^{3}).
\end{split}
\end{align}
In Fig.~\ref{fig:sup_3} we plot the spectrum of the effective non‑Hermitian Hamiltonian obtained by numerical diagonalization in comparison with Eq.~(\ref{Energy}). One can see that the two spectra are very close to each other for $\gamma < J$.

It is instructive to analyze the inverse life-time $\Gamma_{p}$ defined through ${\rm Im} (E_{p}) = -\Gamma_{p}/2$. To first order in $\gamma$ we have
\begin{equation}\label{Gamma1_p}
\Gamma_{p} = \frac{4\gamma}{L+1}\,\sin^{2}\!\left(\frac{\pi p}{L+1}\right).
\end{equation}
The most long‑lived resonances correspond to $p = 1,L$. In the limit of large $L$
\begin{equation}\label{Gamma1}
\Gamma_{\rm max} \approx \frac{4\gamma\pi^{2}}{(L+1)^{3}}.
\end{equation}
Hence, the relaxation time $T_{\rm r} \sim \Gamma_{\rm max}^{-1}$ scales as
\begin{equation}
T_{\rm r} \sim L^{3},
\end{equation}
which severely limits pseudo-classical simulations with long chains. We note in passing that in the middle of the spectrum the inverse life-time scales as
\begin{equation}
\Gamma_{[L/2]} \sim \frac{1}{L}.
\end{equation}

\end{document}